\documentclass[%
twocolumn,
superscriptaddress,
amsmath,amssymb,
aps,
prl,
]{revtex4-2}
\usepackage[per-mode=symbol]{siunitx}
\usepackage{amssymb}
\usepackage{caption}
\usepackage{ragged2e}
\usepackage{amsmath,amssymb}
\usepackage{float}
\usepackage{xcolor}
\usepackage{braket}
\usepackage{makecell}
\usepackage{booktabs}
\usepackage{graphicx}
\usepackage{dcolumn}
\usepackage{bm}
\usepackage[colorlinks=true,linkcolor=blue,citecolor=blue,urlcolor=black]{hyperref}

\begin{document}

\preprint{}

\title{Coherent Dark State Formation of a Lead-Vacancy Spin Qubit in Diamond
}

\author{Yiyang Chen}
\affiliation{Department of Electrical and Electronic Engineering, School of Engineering, Institute of Science Tokyo, Meguro, Tokyo 152-8552, Japan}
\author{Koyo Hirai}
\affiliation{Department of Electrical and Electronic Engineering, School of Engineering, Institute of Science Tokyo, Meguro, Tokyo 152-8552, Japan}
\author{Tzyy Zheng Neo}
\affiliation{Department of Electrical and Electronic Engineering, School of Engineering, Institute of Science Tokyo, Meguro, Tokyo 152-8552, Japan}
\author{Eiki Ota}
\affiliation{Department of Electrical and Electronic Engineering, School of Engineering, Institute of Science Tokyo, Meguro, Tokyo 152-8552, Japan}
\author{Takashi Taniguchi}
\affiliation{Research Center for Materials Nanoarchitectonics, National Institute for Materials Science, Tsukuba, Ibaraki 305-0044, Japan}
\author{Masashi Miyakawa}
\affiliation{Research Center for Materials Nanoarchitectonics, National Institute for Materials Science, Tsukuba, Ibaraki 305-0044, Japan}
\author{Shinobu Onoda}
\affiliation{Takasaki Advanced Radiation Research Institute, National Institutes for Quantum Science and Technology, 1233 Watanuki, Takasaki, Gunma 370-1292, Japan}
\author{Toshiharu Makino}
\affiliation{Advanced Power Electronics Research Center, National Institute of Advanced Industrial Science and Technology, Tsukuba, Ibaraki 305-8568, Japan}
\author{Mutsuko Hatano}
\affiliation{Department of Electrical and Electronic Engineering, School of Engineering, Institute of Science Tokyo, Meguro, Tokyo 152-8552, Japan}
\author{Takayuki Iwasaki}
\email{Corresponding author: iwasaki.t.c5b4@m.isct.ac.jp}
\affiliation{Department of Electrical and Electronic Engineering, School of Engineering, Institute of Science Tokyo, Meguro, Tokyo 152-8552, Japan}

\begin{abstract}
A lead-vacancy (PbV) center in diamond exhibits coherent emission above the liquid helium temperature, making it highly attractive for quantum network applications. Here, we report the magneto-optical and spin properties of PbV centers in diamond. We record a spin lifetime of \qty{12}{ms} at \qty{7.5}{K} under large off-axis magnetic field. Furthermore, we observe formation of the coherent dark state by coherent population trapping and estimate a spin dephasing time of \qty{177}{ns} at \qty{6.5}{K}. This work demonstrates the outstanding thermal robustness of the PbV spin compared to other group-IV centers above \qty{4}{K}. 
\end{abstract}

\maketitle

The realization of quantum networks is fundamental to the advancement of quantum information processing\cite{Kimble2008}. Beyond establishing a platform for unconditionally secure quantum communication, quantum networks enhance the security of quantum computation and push the sensitivity of quantum metrology toward its physical limits\cite{Awschalom2018}. As such, the development of quantum networks has emerged as a main focus of quantum information processing. Quantum emitters in quantum network nodes including quantum repeaters are essential to achieve scalable quantum networks. Defects in diamond have emerged as promising candidates for the quantum emitters, leveraging their long-lived quantum memory\cite{Yamamoto2026,Sukachev2017,Senkalla2024,Guo2023,Karapatzakis2024} and reliable spin-photon interfaces\cite{Evans2018,Codreanu2025}. The multi-node quantum networks based on nitrogen-vacancy (NV) centers have been demonstrated\cite{Pompili2021}. However, the low concentration into the zero-phonon line (ZPL) and spectral diffusion significantly limit its scalability. To address these challenges, the group-IV defects including silicon-vacancy\cite{Hepp2014}, germanium-vacancy\cite{Iwasaki2015}, tin-vacancy\cite{Iwasaki2017}, and lead-vacancy\cite{Wang2021} (SiV, GeV, SnV, and PbV, respectively) have been proposed. These defects exhibit a ZPL concentration around an order of magnitude higher than that of NV centers. Furthermore, owing to their intrinsic inversion symmetry, spectral diffusion is significantly suppressed. These favorable properties have already enabled milestones such as metropolitan-scale quantum networks\cite{Knaut2024}, blind quantum computing\cite{Wei2025}, and enhanced quantum interferometry\cite{Stas2026} using SiV centers.\par
However, the group-IV centers are susceptible to phonon-mediated decoherence in the ground state. Consequently, the SiV and GeV centers require millikelvin temperatures to obtain a long spin coherence time by suppressing the phonon interaction\cite{Sukachev2017,Senkalla2024}. The SnV center typically necessitates temperatures around \qty{2}{K}\cite{Guo2023} below the liquid helium temperature. In contrast, the PbV center stands out with the largest ground state zero field splitting (ZFS) of $\sim$\qty{3900}{GHz}\cite{Wang2021} among the group-IV centers, which significantly suppresses the electron-phonon interaction. This makes the PbV center a unique candidate predicted to have a long spin coherence time above \qty{4}{K}. The suppression of the phonon interaction in the PbV center has been observed as the transform-limited linewidth even above \qty{10}{K}\cite{Wang2024}. However, the spin properties of the PbV qubit have remained elusive. In this work, we report the magneto-optical and spin properties of the PbV centers. We demonstrate all-optical spin initialization and readout, a long spin relaxation time ($T_{1}$), and the coherent dark state formation leading to the estimation of a spin dephasing time ($T^{*}_{2}$). Our results verify the predicted excellent thermal robustness of PbV over \qty{4}{K}, highlighting its potential for scalable quantum networks.\par
Upon Pb ion implantation into the diamond lattice, followed by high-temperature annealing under high-pressure\cite{Wang2021,Abe2026}, Pb atoms combine with vacancies in diamond to form a split-vacancy structure oriented along the $\langle 111 \rangle$ crystallographic axis. This specific configuration constitutes the PbV color centers in diamond with localized energy levels. The atomic structure and the energy level of the PbV center are shown in Fig.~\hyperlink{fig:1}{1(a)}. The centrosymmetric configuration leads to a small permanent electric dipole moment\cite{Santis2021, Aghaeimeibodi2021}, making it highly robust for the first-order electric field fluctuation. Thus, we can obtain stable and narrow emission over time under resonant excitation\cite{Wang2024}. The PbV center is a spin-1/2 system with split ground and excited states. In the absence of a magnetic field, the PbV center has four possible optical transitions, in which only C and D transitions can be observed at low temperatures\cite{Wang2021}. The C and D transitions have wavelengths of approximately 550 and \qty{554}{nm}, respectively, at cryogenic temperatures (Fig.~\hyperlink{fig:1}{1(b)}). The energy difference between the C and D peaks corresponds to a ZFS of $\sim$\qty{3900}{GHz} in the ground state, which is the largest among the group-IV vacancy centers in diamond. This huge ZFS effectively suppresses the phonon-mediated spin relaxation. Consequently, it is predicted the PbV center has the capability to function as a spin qubit at temperatures above the liquid helium temperature\cite{Wang2021,Ruf2021}.\par
The application of an external magnetic field lifts the Kramers degeneracy, thereby enabling us to access the spin degrees of freedom, generating four possible spin-related optical transitions referred to A1, A2, B1, and B2 in Fig.~\hyperlink{fig:1}{1(a)}. The A1 and B2 transitions are spin-conserving, while the other two transitions are spin-flipping. We perform photoluminescence excitation (PLE) spectroscopy to observe the spin dependent optical transitions (Fig.~\hyperlink{fig:1}{1(c)}). In this measurement, we employ a cryostat equipped with a superconducting magnet to apply variable magnetic fields along the $[001]$ direction of the diamond sample. In this configuration, the angle between the magnetic field and the quantization axis of PbV is approximately \ang{54.7}. To prevent the PLE spectra from being invisible due to a long $T_{1}$ time \cite{Sukachev2017,Senkalla2024}, this experiment is conducted at \qty{13}{K}. At zero field, we observe a sharp resonance with a linewidth of \qty{38}{MHz} derived from a single Lorentzian fitting, which corresponds to the transform-limited linewidth of the PbV center\cite{Wang2024}. Similar to our previous work \cite{Wang2024}, the observation of the transform-limited linewidth above \qty{10}{K} demonstrates the exceptional optical coherence of the PbV center benefiting from its large ZFS in the ground state. As the magnetic field strength increases, the two spin-conserving transitions become distinguishable. The split width of the spin-conserving transitions with respect to the magnetic field varies by \qty{5.98}{GHz/T}. Note that although the magnetic field misalignment introduces a spin mixing\cite{Hepp2014,Thiering2018}, it is thought that the mixing degree is not sufficient for the direct observation of the spin-flipping transitions. \par

\begin{figure}[htbp]
  
  \hypertarget{fig:1}{\includegraphics[width=3.375in]{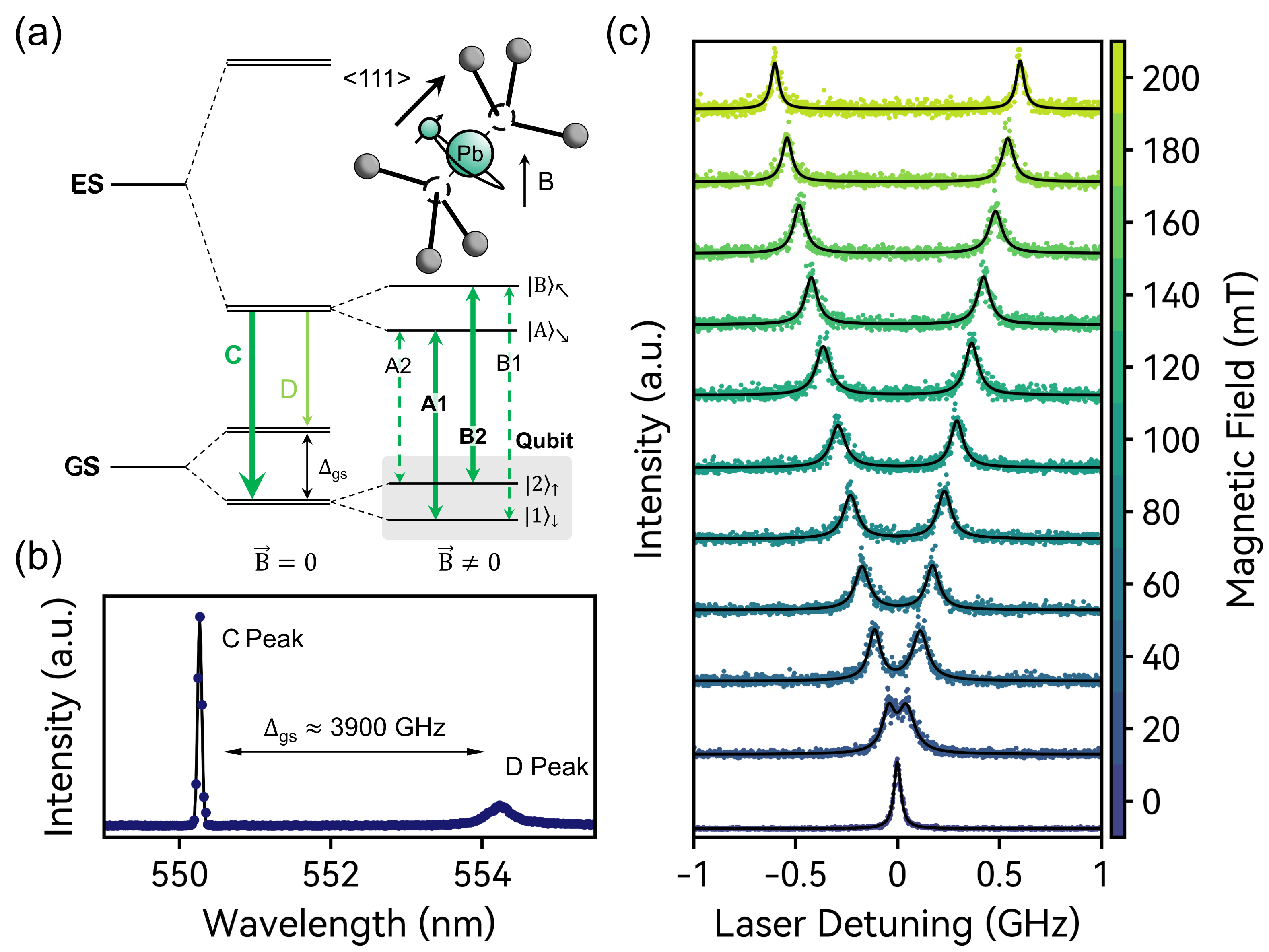}}
  \caption{Magneto-optical properties of the PbV center. (a) Atomic structure and energy level of the PbV center in diamond. (b) Cryogenic photoluminescence (PL) spectrum under non-resonant \qty{532}{\nm} laser excitation. (c) PLE spectra under various magnetic fields at \qty{13}{K}. The color corresponds to the magnetic field strength, varied from 0 to \qty{200}{\mT} in \qty{20}{\mT} increment. The solid lines represent fits using Lorentzian functions.}
\end{figure}

Optically addressing the spin degrees of freedom enables us to perform time-resolved measurements to characterize the dynamic spin properties in the system. The following measurements are performed on a PbV center in solid immersion lens (SIL) to enhance the photon extraction efficiency\cite{Hirai2026}. An inset of Fig.~\hyperlink{fig:2}{2(a)} shows a confocal fluorescence mapping (CFM) of the target SIL, where stronger photon emission is observed from the lens with a diameter of approximately \qty{1}{\um}. In subsequent experiments, we utilize a neodymium magnet to apply a static magnetic field to the PbV center in another cryostat. A magnetic field with a strength of approximately \qty{220}{\mT} is applied along the $[001]$ direction to the sample.\par

We investigate the spin state initialization which occurs through the spin-flipping channel from the excited state. Following a \qty{532}{\nm} charge state initialization pulse with \qty{5}{\ms} duration\cite{Chen2025}, a resonant excitation pulse to the spin-conserving transition B2 is applied and the fluorescence in the phonon sideband (PSB) is collected. We observe a transient decay in the fluorescence intensity (Fig.~\hyperlink{fig:2}{2(a)}). The first non-spin-selective \qty{532}{\nm} laser irradiation establishes a balanced spin population in both the $\ket{1}_\downarrow$ and $\ket{2}_\uparrow$ sublevels. The following spin-selective resonant excitation pumps the population into the opposite spin state via the excited state. This process is attributed to the existence of the spin-flipping relaxation channel in the excited state, which is facilitated by the combined effects of magnetic field misalignment and Jahn-Teller effect or strain environment. Consequently, the spin state of the PbV qubit is optically initialized. The contrast between the initial fluorescence intensity and well-initialized steady state yields a spin initialization fidelity of 98.7\%. \par

Then, we investigate the laser power dependence of the spin initialization rate (Fig.~\hyperlink{fig:2}{2(b)}). This dependence provides insight into the branching ratio (cyclicity) $\eta$ – the ratio between the spin-conserving and spin-flipping relaxation channels from the excited state. The experimental data with a saturation behavior are fitted by $\frac{\Gamma}{2}\frac{P/P_{\text{sat}}}{1+P/P_{\text{sat}}}\frac{1}{\eta}$ \cite{Debroux2021}, where $\Gamma$ is the spontaneous emission rate derived from the excited state lifetime ($\Gamma=1/\tau$). The saturation power $P_{\text{sat}}$ and branching ratio $\eta$ are fitting parameters. For the PbV center in the SIL, we extract a $P_{\text{sat}}$ of \qty{3.1}{nW} and a branching ratio of 550. The branching ratio obtained here is comparable to that of SnV center under a similar magnetic field configuration\cite{Gorlitz2022}. \par

The branching ratio is fundamentally linked to the single-shot readout (SSR) fidelity. A high branching ratio indicates that spin-conserving relaxation channel through the excited state is predominant, allowing for an extended readout time\cite{Sukachev2017,Chen2022,Rosenthal2024} and, consequently, a higher number of detected photons. Conversely, rapid spin-flipping relaxation channel to the opposite spin state results in poor photon detection and degrade the SSR performance. We use the sequence shown in Fig.~\hyperlink{fig:2}{2(c)} to evaluate SSR of the PbV center\cite{Gorlitz2022,Gundlapalli2025}. To separately resonantly excite the spin-conserving A1 and B2 transitions, we use two tunable lasers that lock to the A1 and B2 transitions by PID loops from a wavemeter. Following a repump pulse of \qty{532}{\nm} laser irradiation, a \qty{300}{\us} pulse is applied to excite the B2 transition. This process initializes the system into the $\ket{1}_\downarrow$ spin state, resulting in the nearly entire electron population residing in $\ket{1}_\downarrow$ spin sublevel state. Following a \qty{10}{\us} time delay, a \qty{300}{\us} readout pulse resonantly excites the A1 transition, during which the continuous photon emission is accompanied by the simultaneous pumping of the electron into the opposite spin sublevel. Eventually, the population in the $\ket{1}_\downarrow$ sublevel is almost depleted. After \qty{10}{\us} idle time, a resonant pulse (referred to dark pulse) to the A1 transition is applied again. Due to the absence of the electron population in this sublevel, only few photons could be detected. We repeat this sequence 10,000 times, recording the photon number detected in each readout and dark pulse. Although photon arrival events are recorded over a total duration of \qty{300}{\us}, time-resolved analysis indicates a optimal readout window of \qty{22}{\us}. The photon number statistics under optimal readout window are shown in Fig.~\hyperlink{fig:2}{2(d)}. Benefiting from the enhanced photon extraction efficiency with the SIL structure, we detect 3.23 photons on average for the readout pulse in the optimal readout window. In contrast, the average count  in the optimal readout window of the dark pulses is 0.05. We set one photon as a discrimination threshold between the bright and dark states. The SSR fidelity is defined as $\mathcal{F}_{SSR}=1-\mathcal{E}$\cite{Myerson2008}, where $\mathcal{E}=\frac{1}{2}\left(\mathcal{E}_\text{R}+\mathcal{E}_\text{D}\right)$. The $\mathcal{E}_\text{R}$ ($\mathcal{E}_\text{D}$) is the possibility of obtaining photon counts below (above) the threshold during a single readout (dark) pulse. We achieve a SSR fidelity of 81\%, which is comparable to those of the GeV\cite{Chen2022} and SnV\cite{Gorlitz2022} under the similar magnetic field configuration. This value surpasses the 50\% threshold of random guessing, demonstrating the feasibility of SSR for the PbV center even placed in a large off-axis magnetic field environment. Furthermore, precisely aligning the magnetic field could increase the branching ratio, thereby yielding a higher photon count during the readout and improving the SSR fidelity. \par

\begin{figure}[htbp]
  
  \hypertarget{fig:2}{\includegraphics[width=3.375in]{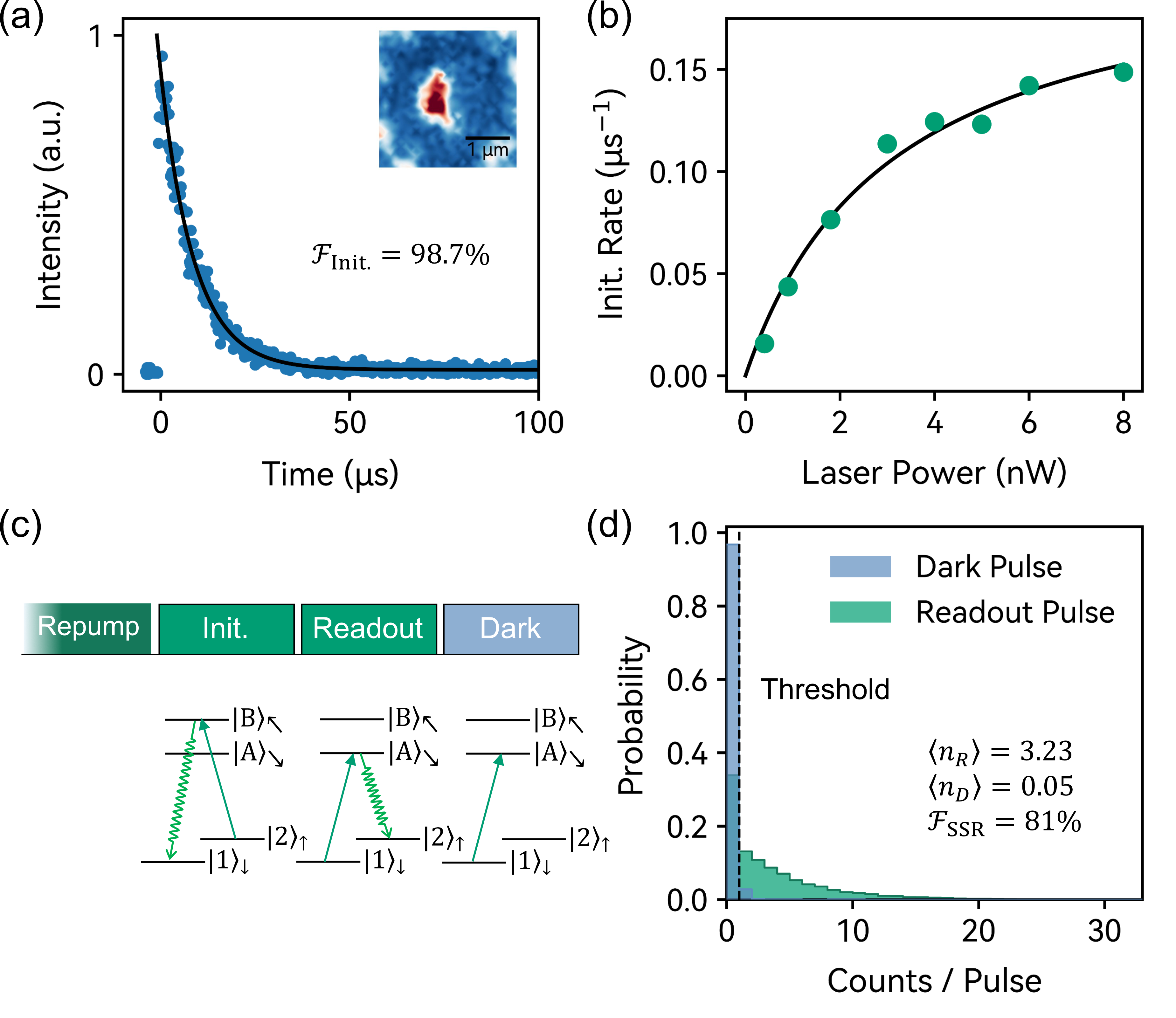}}
  \caption{Spin initialization and SSR performance of the PbV center at \qty{7.5}{K} (a) Spin initialization. Inset shows a CFM image of the PbV in SIL. (b) Spin initialization rate as a function of the resonant laser power. Solid line denotes the fitting described in the main text. (c) Pulse sequence implemented for the SSR measurement. (d) Photon counting statistics derived from the sequence in (c) under optimized readout window. The threshold is set to one photon per pulse to discriminate the bright and dark states, yielding a SSR fidelity of 81\%}.
\end{figure}

On the basis of the spin state initialization and readout above, we evaluate longitudinal spin relaxation time ($T_{1}$) of the PbV center. The upper panel in Fig.~\hyperlink{fig:3}{3(a)} depicts the measurement sequence. After the charge state stabilization pulse, the initial spin-selective B2 excitation polarizes the electron population into the $\ket{1}_\downarrow$ spin sublevel. A varied time delay is introduced to allow the phonon-mediated relaxation. Finally, a B2 pulse is applied to probe the recovered population into the $\ket{2}_\uparrow$ state, thereby quantifying the spin-flip probability induced by the phonon interaction during the delay time. Here, photons in a \qty{10}{\us} collection window are extracted for data analysis. As shown in Fig.~\hyperlink{fig:3}{3(a)}, the fluorescence counts gradually recover toward saturation as the delay time increases, indicating the relaxation of the spin population back to the thermal equilibrium via the electron-phonon interaction. We fit the result with a mono-exponential function, resulting in $T_{1}=\qty{12(2)}{\ms}$ at \qty{7.5}{K}, which is the highest value under similar magnetic field configuration at this temperature in the group-IV vacancy centers in diamond\cite{Rogers2014,Becker2018,Pingault2017,Siyushev2017,Trusheim2020}. For $T_{1}$ over \qty{10}{\ms} under similar magnetic field configuration, the SnV center should be below the liquid helium temperature regime \cite{Trusheim2020}. This fact confirms that the electron-phonon interaction is largely suppressed by the large ZFS in the ground state for the PbV center.\par
 Figure~\hyperlink{fig:3}{3(b)} shows the temperature dependence of $T_{1}$. As a function of inverse temperature on a semi-logarithmic scale. We first attempt to describe the temperature dependence of $T_{1}$ using a modified Orbach process defined by the function $\frac{1}{T_1} \propto \frac{\Delta_{\text{gs}}^3}{\exp\left( \frac{\alpha h \Delta_{\text{gs}}}{k_{\text{B}} T} \right) - 1}$\cite{Rosenthal2023}, where $\Delta_{\text{gs}}$ is a ground state splitting of \qty{3903}{\GHz} for the emitter used here and $\alpha$ is a scaling factor. The fitting to this function, shown as a dashed line in Fig.~\hyperlink{fig:3}{3(b)}, yields a scaling factor of 0.5. However, the Orbach process\cite{Jahnke2015} is a fundamentally resonant two-phonon process that requires a real additional energy level, which is the upper branch level for the group-IV vacancy centers. Thus, the obtained scaling factor of 0.5 is inconsistent with this physical model. A possible explanation is the additional contribution from the Raman process. A model incorporating a Raman term ($1/T_1\propto T^7$)\cite{Wolfowicz2021} with the Orbach process with $\alpha$=1 fits the results well (solid line in Fig.~\hyperlink{fig:3}{3(b)}). Therefore, we suggest that the spin-lattice relaxation of PbV at this temperature regime is governed adjointly by the Orbach and Raman processes.\par

\begin{figure}[htbp]
  
  \hypertarget{fig:3}{\includegraphics[width=3.375in]{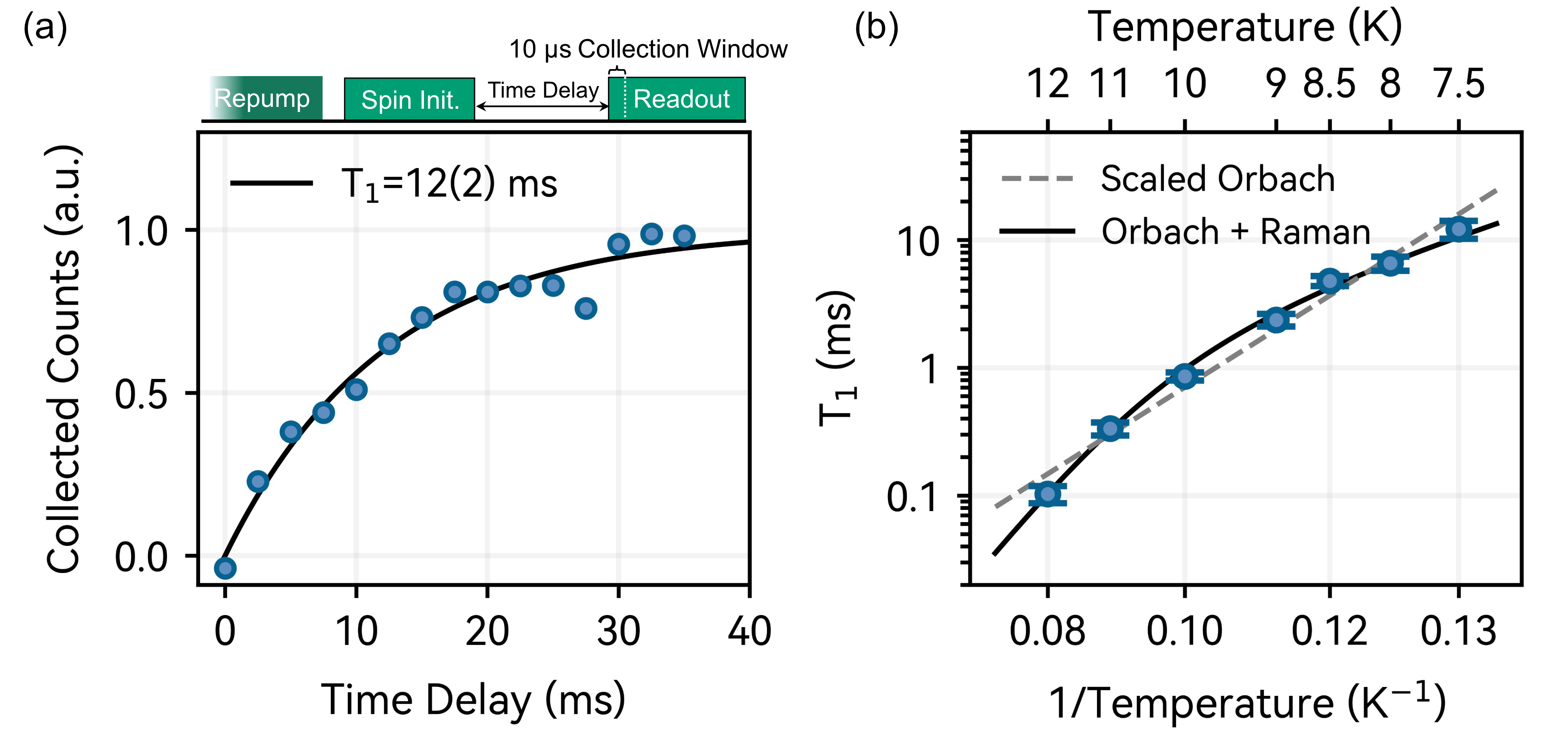}}
  \caption{Longitudinal relaxation time.
  (a) $T_1$ measurement at \qty{7.5}{K}. (b) Temperature dependence of $T_1$.}
\end{figure}

Finally, we observe the coherent dark state of the PbV spin via coherent population trapping (CPT). CPT is an all-optical method to probe the phase coherence of a spin system by using a shared excited state\cite{Pingault2014,Rogers2014}. As shown in an inset of Fig.~\hyperlink{fig:4}{4(a)}, a pump laser is locked to a spin-conserving transition of B2, while a second probe laser is scanned in the frequency domain relative to the pump laser. The frequency offset between the two lasers is defined as the Raman frequency. When the probe laser satisfies the zero two-photon detuning condition, i.e., becoming resonant to the shared excited state, a $\Lambda$ scheme is established. In this scheme, destructive quantum interference traps the electron population in a non-interacting superposition dark state, resulting in the observation of a dip in spectrum. Here, we implement a phase modulation electro-optic modulator to realize high mutual optical coherence of the pump and probe lasers during the Raman frequency scanning. The sequence used for CPT is shown at the upper panel of Fig.~\hyperlink{fig:4}{4(a)}. Following charge state stabilization pulse, the electron population is initialized to $\ket{1}_\downarrow$. Then, the pump and probe lasers are simultaneously irradiated to the PbV center. This sequence is repeated for different Raman frequencies. In a representative CPT curve in Fig.~\hyperlink{fig:4}{4(a)}, a high-contrast fluorescence dip is clearly observed, signifying the formation of a coherent dark state. A black solid line is the model fit based on the Lindblad master equation\cite{Lindblad1976}. The frequency of the resonance dip corresponds to a qubit frequency of \qty{4.24}{\GHz} between the sublevels $\ket{1}_\downarrow$ and $\ket{2}_\uparrow$. The intrinsic linewidth of a CPT dip is fundamentally determined by the spin dephasing time ($T^{*}_{2}$). To eliminate the effect of power broadening, we perform a power dependent measurement and extract the linewidths of the CPT dip by Lorentzian function fitting. Figure~\hyperlink{fig:4}{4(b)} shows that the CPT linewidth measured at \qty{6.5}{K} becomes narrower as the laser power is reduced. We summarize the CPT linewidth as a function of the laser power in Fig.~\hyperlink{fig:4}{4(c)}. At the minimum power limit, we observe the narrowest linewidth of \qty{1.8(13)}{\MHz}, corresponding to a spin dephasing time of \qty{177}{\ns}. This extracted $T_{2}^{*}$ is close to the spin-bath limit for Group-IV centers in diamond\cite{Sukachev2017,Trusheim2020,Rosenthal2023,Sohn2018}.  The relatively large estimation error may stem from the insufficient photon counts in the low drive power regime of the CPT measurements. Therefore, direct measurements of $T_2^*$ using Ramsey sequences\cite{Sukachev2017} in future work will provide higher accuracy.

\begin{figure}[htbp]
  
  \hypertarget{fig:4}{\includegraphics[width=3.375in]{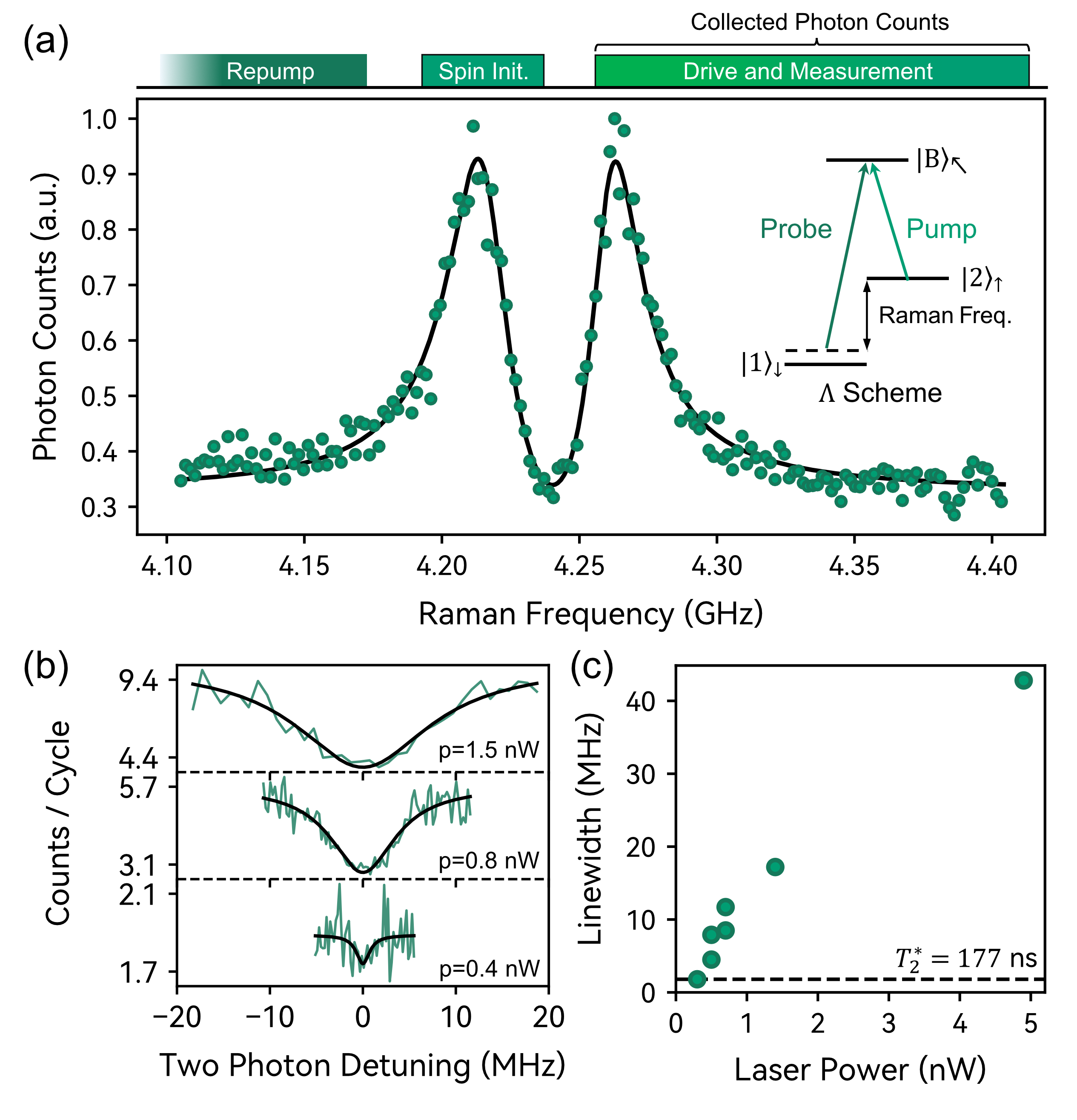}}
  \caption{CPT of the PbV center. (a) Sequence and a representative CPT spectrum at \qty{7}{K}. The black solid line represents a numerical fit based on the Lindblad master equation. (b) Evolution of the CPT spectra as varying the laser power at \qty{6.5}{K}. (c) CPT linewidth  as a function of the total incident laser power at \qty{6.5}{K}. The narrowest observed linewidth is \qty{1.8(13)}{\MHz}, corresponding to a spin dephasing time ($T^{*}_{2}$) of \qty{177}{\ns}}.
  
\end{figure}

The evaluation of $T_{1}$ underscored the extraordinary potential of the PbV center for the operation above the liquid helium temperature. Under a significant off-axis magnetic field, we recorded $T_{1}$ of \qty{12}{ms} at \qty{7.5}{K}. This performance significantly shifts the temperature benchmark for achieving $T_{1}>\qty{10}{ms}$ compared to the \qty{3.25}{K} reported for the SnV center\cite{Trusheim2020} under similar magnetic field condition. Also, the spin dephasing time recorded at \qty{6.5}{K} approaches the spin-bath limit, indicating that phonon-mediated decoherence is mitigated. A $^{\text{12}}\text{C}$ enriched diamond could decrease the spin bath noise\cite{Sukachev2017,Guo2023,Balasubramanian2009}, thereby significantly reducing the phase noise and extending the spin dephasing time. Such an isotopically purified environment also creates a favorable condition for achieving a long spin coherence time. According to the prior study on SnV\cite{Guo2023}, the $T_{1}$ of PbV centers can be enhanced by several orders of magnitude through precise magnetic field alignment to the symmetry axis, suggesting that $T_{1}$ potentially exceeds \qty{1}{\s} at temperature well above \qty{4}{K}. Therefore, we have a reasonable expectation that at elevated temperature, the PbV center can achieve a long spin coherence time that are comparable to those of SiV and GeV at dilution refrigerator temperatures and SnV centers at temperature around \qty{2}{K}.\par
In this work, we report the spin properties of the PbV center in diamond using all-optical methods. We demonstrated the feasibility of SSR, long spin lifetimes, and spin dephasing time of the PbV center above \qty{4}{K}. The excellent thermal robustness of PbV over \qty{4}{K} not only allows for significantly relaxed cryogenic equipment but also enables the use of a higher microwave power for the faster gate operation under the microwave spin control scheme in the future, which makes the PbV center as an attractive candidate for nodes in scalable quantum networks.\par

\vspace{\baselineskip}

The authors would like to thank Ryotaro Abe and Keita Ikeda for experimental support. We thank Materials Analysis Division, Core Facility Center, Institute of Science Tokyo for technical assistance. This work is supported by JSPS KAKENHI Grant Number 25K24501, the MEXT Quantum Leap Flagship Program (MEXT Q-LEAP) Grant Number JPMXS0118067395, JST Moonshot R\&D Grant Number JPMJMS2062, JST ASPIRE JPMJAP24C1, JST SPRING Grant Number JPMJSP2180, and Council for Science, Technology and Innovation (CSTI), 3rd Cross-ministerial Strategic Innovation Promotion Program (SIP) Quantum.

\bibliography{apssamp}

\end{document}